# RECOMMENDATION FOR WEB SERVICE COMPOSITION BY MINING USAGE LOGS

Vivek R, Prasad Mirje and Sushmitha N

Department of Information Science, RVCE, Bangalore, India

## ABSTRACT

*Web service composition has been one of the most researched topics of the past decade. Novel methods of web service composition are being proposed in the literature include Semantics-based composition, WSDL-based composition. Although these methods provide promising results for composition, search and discovery of web service based on QoS parameter of network and semantics or ontology associated with WSDL, they do not address composition based on usage of web service. Web Service usage logs capture time series data of web service invocation by business objects, which innately captures patterns or workflows associated with business operations. Web service composition based on such patterns and workflows can greatly streamline the business operations. In this research work, we try to explore and implement methods of mining web service usage logs. Main objectives include Identifying usage association of services. Linking one service invocation with other, Evaluation of the causal relationship between associations of services.*

## 1. INTRODUCTION

Nowadays a significant part of a software system is structured using software services and implemented using Web service technologies. Web services have become standard, collaborative and can handle a complex conversation with other software and service systems. Service composition is implementation technology that defines how to implement a web service by composing other services. Web service composition is essentially operation composition provided by web service. Business process architects discover [10] required web service to achieve a given goal and compose them into an application using specification language or service composition language (e.g., XL, WSFL, and BPML [10]). The main challenges in service composition are a lack of domain knowledge about a business process which is being implemented and identify a set of service that can be used to execute a business process [11].

To address aforementioned challenges many types of researches have been carried out, they are divided into two paradigms, one where we define business process first (goal achieved by composed service) and then identify set of service which satisfies the business process, this can paradigm can be regarded as business process oriented service composition (BPOSC). The other approach would be to identify set of service which are frequently being invoked, associate them and identify a workflow from the association and evaluation of workflow in terms of validity and commercial benefits, such composition paradigm can be regarded as invocation or Execution based service composition (EBSC). Recommendation systems for BPOSC aid the user in identification of business process, defining the business process and searching of services which satisfies the business process, to this regard many types of research have been carried out, QoS based searching [9], and syntactic & semantic based searching methods have been developed.

Server logs are rich input to provide recommendations for EBSC, many data mining techniques like association rule mining, clustering can be easily adopted which can aid the user in associating web services, identifying workflows and validation of workflows

In this research work, we try to explore and implement methods of mining web service invocation logs. Main objectives include Identifying usage association of services. Linking one service invocation with other. We will try to evaluate popular data mining techniques which can be used to discover association among web service invocation and visualize these association as a recommendation for service composition.

The remainder of this paper is organized as follows. Section 2 discusses the related work. Section 3 gives an overview of the dataset required for such recommendation and challenges and approach followed in generation of same. Section 4 evaluates the various association mining algorithm and gives details of our approach. Section 5 discusses the results of the approach followed; Section 6 concludes the paper.

## 2. RELATED WORK

There are many approaches to service composition and each approach addresses a different set of challenges of service composition. These approaches can be classified into four different groups [12] process based, model based, logic or calculi based, and AI-based approaches. As described in the previous section this classification applies to business process oriented service composition. Our work presented here relates to execution based service composition approaches. The concept of application of data mining is explored in [4], recommends a set of applications that can leverage problems concerned with the planning, development and maintenance of Web services and their compositions. [14] Presents a similar approach of applying Web service log-based analysis and process mining techniques in order to provide a mean to ensure a correct and reliable Web services modelling and execution. They also present the issues and challenges in collecting execution logs of service invocation. [13] Proposes an approach that automatically identifies the service composition patterns from various applications using execution logs. The approach employs Apriori algorithm to mine associated services. Analysing the control flow of services and the order of service invocation based on the execution logs. In our research we go beyond the use of simple Apriori for associating web services, we exploit the natural hierarchy that exists between service and operation, in order to generate better and reliable service association that can be used as a recommendation for service composition.

## 3. DATASET GENERATION

Although weblogs mining is an emerging research topic for recommendation systems, the publically available datasets for researchers are more concentrated towards the recording of invocation of web pages rather than web service. Such weblogs record the hits a web page has received. Available datasets for research under enhancement of web services are either QoS based [6, 7, 8] or WSDL-based [9]. Minimalistic web service invocation logs required should contain parameters like time of invocation, client IP or session identifier and web service URL, additionally the logs may also contain QoS information like data size, response time, HTTP code and message.

Due to lack of such detailed web service invocation logs, we have resorted to the generation of web service logs. This section gives brief explanation of methods followed during data set generation. The generated dataset consists of parameters as shown in Table 3, for simplicity we have considered web service name instead of web service URL. Usually, a single web service exposes multiple operations, which is indicated by operation column. In a composite web service, the order in which these web services are invoked are also important, this order of invocation is governed by WS coordination protocol. The recommender system should maintain this sequence of invocation.

Table 1: Generated dataset

| Session id | Time stamp | Web service name | Operation | Response time | Response size |
|---|---|---|---|---|---|
| ... | … | … | … | …. | …. |

We consider a set of 50 compositions consisting of 100 web service (WS1, WS2…WS100), each web service exposes randomly chosen number of operations ranging from 2 to 15 as our final result to be produced. Using the same 100 web services, we generate invocation logs, the generated logs consist of 1000 session. In each session, a randomly chosen web service will be invoked. The order of invocation is also random. One may argue that the occurrence of any of the composition in the above-said list would be very sparse due to randomness involved in the generation of logs. When any mining technique applied on such sparse data, a threshold for interestingness criteria (like support and confidence under association rules) which would help us recommend service composition should be set to a lower value. This would lead in increased number of candidate compositions, and also reduced the performance of the mining algorithm. Hence, for every randomly chosen invocation between range of 2 to 10, one of the compositions in the list will be randomly chosen and included as subsequence in the invocation logs. This would bring a balance between randomness and sparse-city in the data and generate fair web service invocation logs, which can be used to evaluate different mining procedure best suited for recommender system of WS compositions.

## 4. PROPOSED METHODOLOGY

Table 2: Performance of various algorithm

| Algorithm | Min support (%) | Min confidence (%) | Candidate sequence count | Frequent sequence count | Number of association rules |
|---|---|---|---|---|---|
| Apriori | 3 | 3 | 149490 | 115001 | 25366354 |
| Apriori | 3.5 | 3.5 | 53038 | 49624 | 12664936 |
| Apriori | 3.5 | 3.8 | 49624 | 49624 | 12664936 |
| Apriori | 3.7 | 3.7 | 87 | 62 | 360 |
| Apriori | 3.8 | 3.8 | 63 | 63 | 602 |
| Closed association rules | 3.7 | 3.7 | 2 | 2 | 60 |
| FP growth | 3.7 | 3.7 | 62 | 62 | 360 |

Table 2: shows performance of various well-known data mining techniques that can be used for generation of associations of web composition. A set of 1000 sessions consisting of service invocation was used as input. We can observe that when the overall value of mini support & min confidence is low, and chosen different value either generates too many candidates or too few of them, which in turn results in varying number of association rules. We can see that the efficiency of such algorithm greatly depends on the interestingness criteria given as input (support and confidence).

The observed efficiency can be greatly increased by modelling some application specific knowledge into the system. Service specific information like description of service [1] in WSDL, QoS [6, 7] provided by service, invocation order provided coordination protocol can be used. Use

85

of service description and QoS parameter have been greatly explored in the literature. In this paper, we will explore the invocation order as a parameter that can benefit the recommender system. Invocation order can either be derived from coordination protocol or discovered from invocation logs. Invocation order derived from coordination protocols provide as information regarding service composition in practice, they help in verifying whether implemented service compositions are of value addition or not, rather than helping to discover new service composition. On the other hand discovered order can be used to identify new service composition.

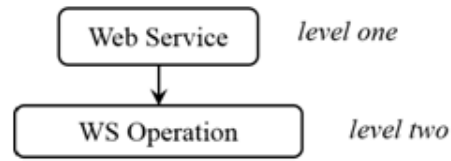

Figure 1: Concept hierarchy for multilevel association rule mining

Figure 1 shows the hierarchal view of the composed list of service, i.e., each web service performs or consists set of operations. The findings of service composition can be done iteratively or step-wise using this hierarchy. This technique has an intrinsic way to reduce candidate list. Learning of one iteration is used to reduce the candidate set to be processed in subsequent phases.

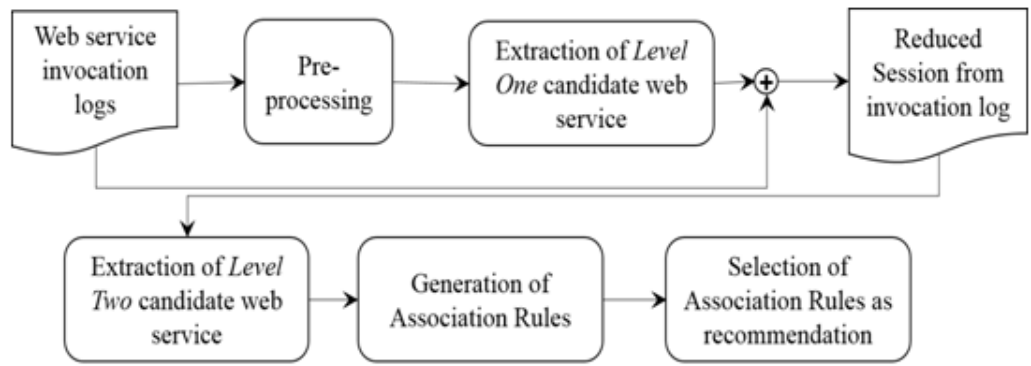

Figure 2: Proposed methodology for web service composition using multi-level association rules

Figure 2 gives the overview of the approach to extract service composition recommendation from execution log. Invocation logs captured by the web server is given as input to the system. Logs are pre-processed as initial step which includes process of identification of sessions, client ids, client groups and transformations making the logs minable by data mining methods. In next step the whole logs are scanned and from identified sessions, web service names are extracted excluding the operations involved in the invocation. If a web service is invoked multiple times with same operations (as in loop invocation) or different operations, single occurrence will be considered, for e.g. consider a session with invocation {A (a), A (b), C (d), D (a), A (c)} (where A, C, D are web service names and a, b, c, d are operations with-in web service) will be transformed to {A, C, D}. Such transformations which contain only web service names forms level one of the concept hierarchy. To preserve the ordering in composed service, invocations in a session are treated as sequence. These candidate sequences can occur as complete sequence or sub-sequence in sessions, the number of occurrence in logs forms the support of candidate sequence. There are many ways to find interestingness of a candidate, support being most widely used, we take average support as the threshold. Each sequence corresponds one or more sessions, only those sessions with candidate sequence or sub-sequence whose support is greater than avg. support will be considered for next level of mining. This greatly reduces the number of candidates to be processed, enhances speed and accuracy of the system. Reduced invocation logs will be generated which will serve as input to level two candidate web service generation. The frequent sequence will be found consisting of service names as well as operation being invoked. These candidate WS sequence can be used for generation of association rules, top n rules will be provided as the recommendation to the user.

## 5. EXPERIMENTS AND RESULTS

As discussed in previous sections, a data set of 1000 session was used as input for generation service composition. As discussed in section 3, the evaluating criteria considered in this work is a set of 50 compositions consisting of 100 web services. Association rules generated from proposed method are compared against these to determine the performance.

Table 3: Experimental results

| Algorithm | Min support (%) | Min confidence (%) | Candidate sequence count | Frequent sequence count | Number of association rules | Number of matching association |
|---|---|---|---|---|---|---|
| Apriori | 3.5 | 3.5 | 53038 | 49624 | 12664936 | 50 |
| Apriori | 3.5 | 3.8 | 49624 | 49624 | 12664936 | 50 |
| Apriori | 3.7 | 3.7 | 87 | 62 | 360 | 16 |
| Proposed Method | 3.5 | 3.5 | 899 | 530 | 1370 | 48 |
| Proposed Method | 4.5 | 4.5 | 121 | 188 | 156 | 42 |
| Proposed Method | 6.5 | 6.5 | 84 | 135 | 101 | 35 |

The correctness of approach proposed can be verified by the use of two metrics [13], precision and Noise ratio. Precision is given by equation 1; Precision obtained should be high, as it measure whether the proposed approach is able to identify the associations present in the dataset (execution logs). Here precision can also be considered as accuracy.

$$precision = \frac{\text{\# matching association}}{\text{\# compositions considered in data set}} \quad \ldots \text{eq (1)}$$

Noise vs expected association rules ratio is given by equation 2; Noise ratio should be low, as it is the measure of ability of proposed approach to identify interesting composition from given execution logs. It also signifies the capacity of proposed approach to reduce the candidate set into an association rule. Ideally the ratio obtained should be 1, i.e., when expected number of association is equal to generated number of association. Less ratio signifies possible loss of data or over reduction, and more ratio signifies introduction of noise in the association rules generated. Noisy set of association rules would not be beneficial for recommender system.

$$noise\ ratio = \frac{\text{\#generated association}}{\text{\#expected association}} \quad \ldots \text{eq (2)}$$

Table 3 shows the experimental results in identifying the service composition from invocation logs. Although the performance of Apriori is more accurate when compared to proposed method, the ratio of noise to expected association rules is significantly high, making Apriori computationally heavy, whereas the proposed method would generated almost accurate result with very much less noise in generated association rules. Figure 3 shows the graphical form of Noise vs Expected association rules. Table 4 provides the results obtained from calculations of the metrics. For calculation, the number of expected association is considered as 50, as it the number of composition with which the data set was generated.

Table 4: Correctness metrics for proposed method

| Algorithm | Number of association rules | Number of matching association | Precision (%) | Noise ratio |
|---|---|---|---|---|
| Apriori | 12664936 | 50 | 100 | 253298.72 |
| Apriori | 12664936 | 50 | 100 | 253298.72 |
| Apriori | 360 | 16 | 32 | 7.2 |
| Proposed Method | 1370 | 48 | 96 | 27.4 |
| Proposed Method | 156 | 42 | 84 | 3.12 |
| Proposed Method | 101 | 35 | 70 | 2.02 |

## 6. CONCLUSION

In this research work, a novel approach for extraction of recommendations for web service composition is proposed. The proposed method employs the intrinsic hierarchy available in web service (i.e., every service is composed of set of operations) to reduce the candidate service sequence to be processed. This yields in associations which has higher support. The proposed work also explores invocation order of web services as criteria for service composition recommendation. From the results shown we can conclude that proposed method would generated much lesser noise association rules, hence providing more interesting recommendation for web service compositions.

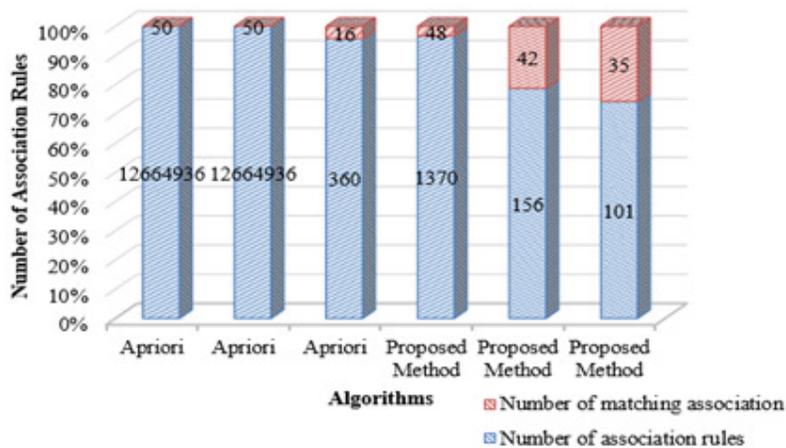

Figure 3: Noise vs Expected Association rules

## REFERENCES

[1] Liang Chen, Jian Wu, Zibin Zheng, Michael R. Lyu, Zhaohui Wu, Modeling and exploiting tag relevance for Web service mining, Knowledge and Information Systems, April 2014, Volume 39, Issue 1, pp 153-173

[2]   Richi Nayak, Aishwarya Bose, A Data Mining Based Method for Discovery of Web Services and their Compositions, Real World Data Mining Applications, Annals of Information Systems Volume 17, 2015, pp 325-342

[3]   Nayak, Richi (2008) "Data mining in web services discovery and monitoring", International Journal of Web Services Research, 5(1). pp. 62-80.

[4]   Nayak, Richi and Tong, Cindy M. (2004) "Applications of data mining in web services", In Proceedings 5th International Conferences on Web Information Systems, Brisbane, Australia

[5]   Bhuvaneswari V and Umajothy P, "Mining Temporal Association Rules from Time Series Microarray Using Apriori Algorithm". In proceedings of Review of Bioinformatics and Biometrics (RBB) Volume 2 Issue 2, June 2013

[6]   Zibin Zheng, Michael R. Lyu, "Collaborative Reliability Prediction for Service-Oriented Systems", in Proceedings of the ACM/IEEE 32nd International Conference on Software Engineering (ICSE2010), Cape Town, South Africa, May 2-8, 2010, pp. 35 – 44.

[7]   Zibin Zheng, Yilei Zhang, and Michael R. Lyu, "Distributed QoS Evaluation for Real-World Web Services," in Proceedings of the 8th International Conference on Web Services (ICWS2010), Miami, Florida, USA, July 5-10, 2010, pp.83-90.

[8]   Yilei Zhang, Zibin Zheng, and Michael R. Lyu, "Exploring Latent Features for Memory-Based QoS Prediction in Cloud Computing," in Proceedings of the 30th IEEE Symposium on Reliable Distributed Systems (SRDS 2011), Madrid, Spain, Oct.4-7, 2011.

[9]   Yilei Zhang, Zibin Zheng, and Michael R. Lyu, "WSExpress: A QoS-aware Search Engine for Web Services," in Proceedings of the 8th International Conference on Web Services (ICWS2010), Miami, Florida, USA, July 5-10, 2010, pp.83-90

[10]  J. Lu, Y. Yu, D. Roy, and D. Saha, "Web Service Composition: A Reality Check", 8th International Conference on Web Information Systems Engineering, Nancy, France, December 3-7, 2007

[11]  Bipin Upadhyaya, Ying Zou1, Shaohua Wang and Joanna Ng, "Automatically Composing Services by Mining Process Knowledge from the Web"

[12]  Baryannis, George, and Dimitris Plexousakis. "Automated Web Service Composition: State of the Art and Research Challenges." ICS-FORTH, Tech. Rep 409 (2010).

[13]  Tang, Ran, and Ying Zou. "An approach for mining web service composition patterns from execution logs." Web Systems Evolution (WSE), 2010 12th IEEE International Symposium on. IEEE, 2010.

[14]  Gaaloul, Walid, Karim Baïna, and Claude Godart. "Log-based mining techniques applied to web service composition reengineering." Service Oriented Computing and Applications 2.2-3 (2008): 93-110